\documentclass[a4paper,12pt,twoside]{article}
\usepackage[left=1.9cm,right=1.9cm,top=2.00cm,bottom=2.54cm]{geometry}
\usepackage[utf8]{inputenc}
\usepackage{bbm}
\usepackage{float}
\usepackage{amssymb}
\usepackage[tbtags]{amsmath}
\usepackage{bm}
\usepackage{xcolor}
\usepackage[titletoc]{appendix}
\usepackage[CJKbookmarks]{hyperref}
\usepackage{bookmark}
\usepackage{graphicx}
\usepackage{caption}
\usepackage{fancyhdr}
\newcommand{\dd}{\text{d}}
\begin{document}
\title{Fluctuations of topological charges in two-dimensional classical Heisenberg model through high-temperature and low-temperature expansions}
\author{Shan-Chang Tang\thanks{Shanghai High School, Shanghai 200231, China}}
\date{}
\raggedbottom
\maketitle

\noindent
ABSTRACT: It is well known that KT transition in 2d XY model is driven by the binding and unbinding of topological defects, which can be characterized by the fluctuation of topological charges inside a region. We extend the idea into the 2d Heisenberg model and calculate the fluctuation through high-temperature and low-temperature expansion respectively. It is found that the fluctuation of topological charges is proportional to the area of the region at high temperatures while obeys the perimeter law at low temperatures.

\section{Introduction}
In 1973, Kosterlitz-Thouless (KT) transition was discovered in two-dimensional XY model~\cite{KosterlitzThouless-1972,KosterlitzThouless-1973,Kosterlitz-1974}. Instead of being driven by fluctuations of order parameters in ordinary condensed matter systems, it is driven by the binding and unbinding of vortices. We have proposed that such a transition can be studied with the fluctuation of topological charges inside a region, which is defined as 
\begin{equation}
    \chi=\langle Q^2 \rangle-\langle Q\rangle^2,\label{eqn:fluc_topo_charge}
\end{equation}
with $Q$ being the topological charges~\cite{Tang&Shi-2025}. In 2d XY model, $Q$ is the number of vortices minus the number of antivortices. If the vortices are unbound at high temperatures, $\chi$ will be proportional to the area of the region. On the other hand, if they are bound at low temperatures, $\chi$ will be proportional to the perimeter of the region.

In 2d Heisenberg model, where the spins possess three components, the situation is much more complicated. Generally speaking, the topological defects in this model are skyrmions~\cite{Skyrme-1962,FertCrosSampaio-2013}, which is conjectured to drive a transion in this model~\cite{BrownCiftan-1993}. However, there are only vortex-like defects at low temperatures~\cite{Klenin-1979,KawabataBishop-1980}, and no complete skyrmions are found in our simulations~\cite{Tang&Shi-2025}. This is because skyrmions are not stable topological defects in 2d Heisenberg model~\cite{Toulouse-1976}. 

Despite the above arguments, the fluctuation of topolotical charges~\eqref{eqn:fluc_topo_charge} can still be utilized to investigate whether there is a transition in 2d Heisenberg model. In the previous article~\cite{Tang&Shi-2025}, we have calculated the fluctuations of topological charges through Monte Carlo simulations. The result is that the fluctuation is proportional to the area at high temperatures while proportional to the perimeter at low temperatures.

In this article, we will calculate the fluctuation through high-temperature expansions and low-temperature expansions theoretically. In the method of the high-temperature expansion, the fluctuation of topological charges is calculated in the orders of $\beta=\frac{1}{k_B T}$ through a diagrammatic technique~\cite{Stanley-1967a,Stanley-1967b,StanleyLee-1971}. In the method of the low-temperature expansion, the fluctuation is calculated in the orders of $k_B T$ through the expansion of the spins around a particular direction~\cite{Zinn-JustinBrezin-1976B,Zinn-JustinBrezin-1976L}.

\section{High-temperature expansion}
In this section, the Hamiltonian for the 2d Heisenberg model is
\begin{equation}
    \mathcal{H}=-J\sum_{\langle ij\rangle}\bm{S}_i \cdot \bm{S}_j,
\end{equation}
where $J$ is the exchange interaction constant and $\bm{S}_i$ represents the normalized spin vector. 
$\langle ij\rangle$ means the nearest-neighbouring summation.

At high temperatures, the average value of a function of spins can be expanded as follows,~\cite{Stanley-1967a}
\begin{equation}
    \langle O[\bm{S}]\rangle=\frac{\text{Tr}\left(O[\bm{S}]e^{-\beta\mathcal{H}}\right)}{\text{Tr}\left(e^{-\beta\mathcal{H}}\right)}
    =\sum_k\frac{(-1)^k}{k!}\alpha_k \beta^k,\label{equ:O_expansion}
\end{equation}
where $\text{Tr}(\cdots)$ means $\int \cdots \int \left[\prod_i\dd\bm{S}_i\delta(\bm{S}_i^2-1)\right]$. 
It can also be transformed to
\begin{equation}
    \text{Tr}\left(O[\bm{S}]e^{-\beta\mathcal{H}}\right)=\sum_k\frac{(-1)^k}{k!}\alpha_k \beta^k\text{Tr}\left(e^{-\beta\mathcal{H}}\right),
\end{equation}
where the factor $e^{-\beta\mathcal{H}}$ can be further expanded, 
\begin{align}
    \begin{split}
        \sum_l\frac{(-\beta)^l}{l!}\text{Tr}\left(O[\bm{S}]\mathcal{H}^l\right)
        =&\left[\sum_k\frac{(-1)^k}{k!}\alpha_k \beta^k\right]
        \left[\sum_m\frac{(-\beta)^m}{m!}\text{Tr}(\mathcal{H}^m)\right]\\
        =&\sum_{k,m}\frac{(-\beta)^{k+m}}{k!m!}\alpha_k\text{Tr}(\mathcal{H}^m)\\
        =&\sum_l\frac{(-\beta)^l}{l!}\sum_{k=0}^{l}\binom{l}{k}\alpha_k\text{Tr}(\mathcal{H}^{l-k}).
    \end{split}
\end{align}
Therefore, 
\begin{equation}
    \text{Tr}\left(O[\bm{S}]\mathcal{H}^l\right)=\sum_{k=0}^{l}\binom{l}{k}\alpha_k\text{Tr}(\mathcal{H}^{l-k})
    =\alpha_l+\sum_{k=0}^{l-1}\binom{l}{k}\alpha_k\text{Tr}(\mathcal{H}^{l-k}).
\end{equation}
So $\alpha_l$ can be calculated hierarchically,
\begin{equation}
    \alpha_l=\nu_l-\sum_{k=0}^{l-1}\binom{l}{k}\alpha_k\mu_{l-k},\label{equ:alpha_l}
\end{equation}
where $\nu_l\equiv\text{Tr}\left(O[\bm{S}]\mathcal{H}^l\right)$ and $\mu_{l-k}\equiv\text{Tr}(\mathcal{H}^{l-k})$.
Since $O[\bm{S}]$ and $\mathcal{H}$ are all functions of $\bm{S}$, calculation of $\text{Tr}(S_x^a S_y^b S_z^c)$ is quite critical.
Actually, it is zero when any of $a,b,c$ is odd. If $a,b,c$ are all even, the trace is (see Appendix \ref{app:Tr_SxSySz})
\begin{equation}
    \text{Tr}\left(S_x^{2k} S_y^{2l} S_z^{2m}\right)
        =\frac{1}{2\pi}\frac{\Gamma\left(k+\frac{1}{2}\right)\Gamma\left(l+\frac{1}{2}\right)\Gamma\left(m+\frac{1}{2}\right)}{\Gamma\left(k+l+m+\frac{3}{2}\right)},
\end{equation}

\begin{figure}[htbp]
    \centering
    \includegraphics[scale=0.5]{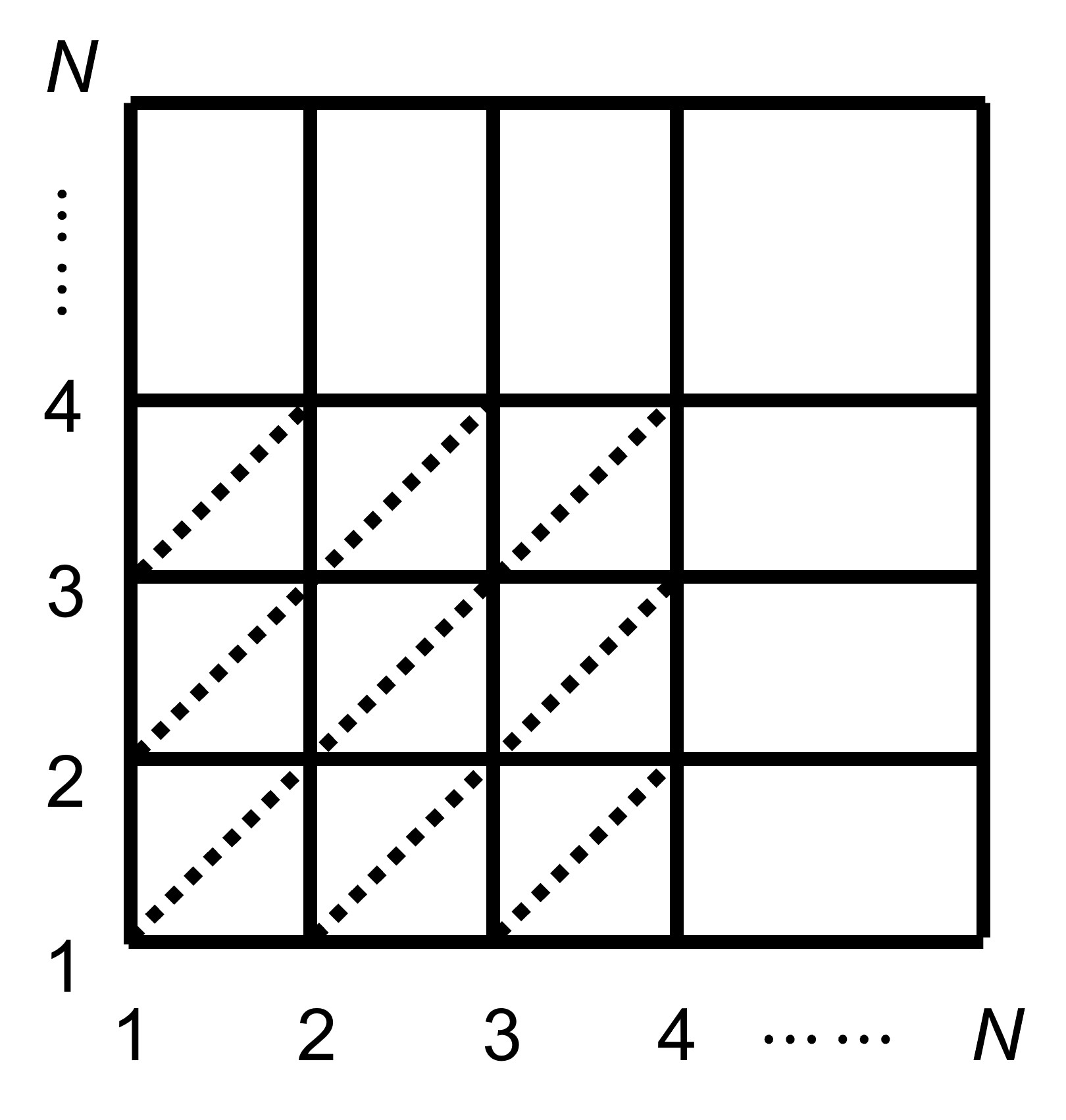}
    \caption{The lattice for calculation}\label{fig:lattice_config}
\end{figure}

In our calculation, the spin on the site of column $i$ and row $j$ is represented by $\bm{S}_{ij}$(See Fig.\ref{fig:lattice_config}).
The topological charge for each plaquette is 
\begin{equation}
    \rho_{ij}\equiv\rho_{ijr}+\rho_{ijl}\equiv\bm{S}_{ij}\cdot(\bm{S}_{i+1,j}\times\bm{S}_{i+1,j+1})+\bm{S}_{ij}\cdot(\bm{S}_{i+1,j+1}\times\bm{S}_{i,j+1}),\label{equ:rho_ij}
\end{equation}
and the topological charge of the whole lattice is $Q=\sum_{ij}\rho_{ij}$.
Similarly, the topological charge of an $L \times L$ square is $Q_L=\sum_{1\leqslant i\leqslant L,1\leqslant j\leqslant L}\rho_{ij}$.

Since
\begin{equation}
    \langle Q_L\rangle=\frac{\text{Tr}\left(Q_L e^{-\beta\mathcal{H}}\right)}{\text{Tr}\left(e^{-\beta\mathcal{H}}\right)}
\end{equation}
contains odd $\bm{S}_{ij}$'s, it must be zero. So the fluctuation of topological charges is
\begin{equation}
    \chi_L=\langle Q_L^2 \rangle=\left\langle\left(\sum_{1\leqslant i\leqslant L,1\leqslant j\leqslant L}\rho_{ij}\right)^2\right\rangle.
\end{equation}
Since the Hamiltonian possesses transitional invariance, the fluctuation can be expanded as
\begin{equation}
    \chi_L=L^2 \langle \rho_{11}^2 \rangle + 4L(L-1)\langle \rho_{11}\rho_{21} \rangle
    + 2(L-1)^2(\langle \rho_{11}\rho_{22} \rangle+ \langle \rho_{12}\rho_{21} \rangle) + \cdots\label{eqn:chi_L}
\end{equation}

\subsection{Calculation of $\mu_i$}
\begin{figure}[htbp]
    \centering
    \includegraphics[scale=0.5]{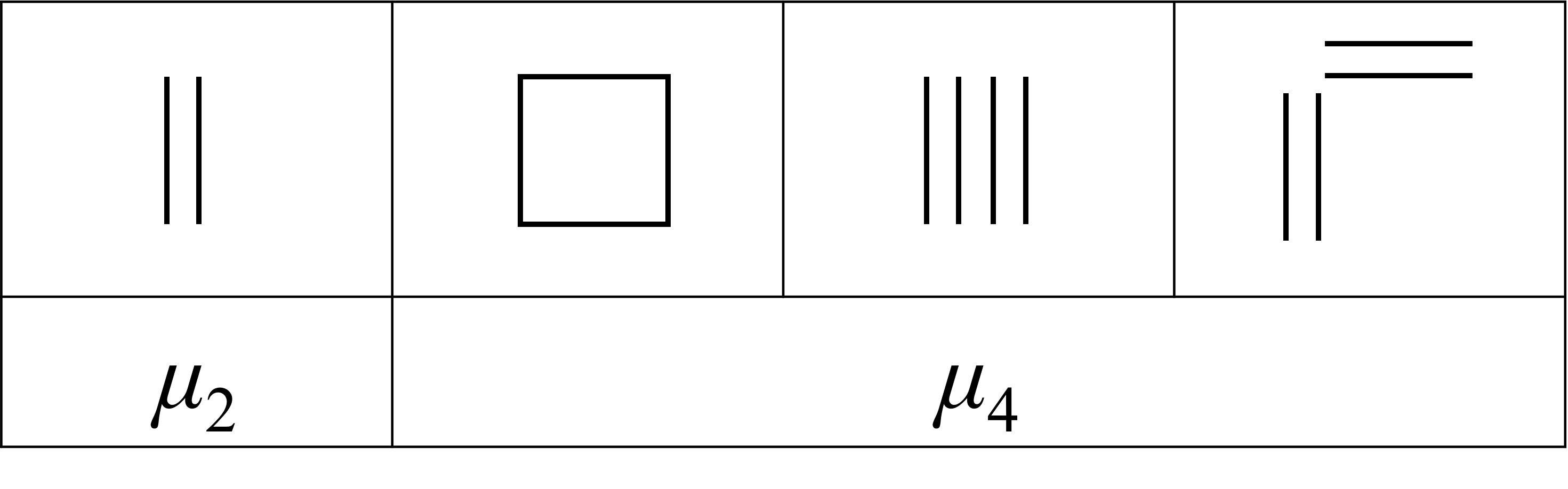}
    \caption{The diagrams for the first several nonzero $\mu_i$'s}\label{fig:mu_i}
\end{figure}

According to the previous studies~\cite{Stanley-1967a,Stanley-1967b,StanleyLee-1971}, $\mu_i$'s can be calculated with the diagrams. 
The components of the Hamiltonian $\bm{S}_i\cdot\bm{S}_j$ are represented by a line linking the site $i$ and the site $j$. 
If the lines form a closed loop on the lattice, every $\bm{S}_i$ will appear twice at least in the trace, so the trace is nonzero. 
Otherwise, the trace will be zero. 
Since the Hamiltonian only includes the nearest-neighbouring interaction, only the nearest-neighbouring sites should be linked by a line.
Another rule is that the diagram should be connected, otherwise it will be cancelled by the $\nu_l$'s.
$\mu_i$'s will be calculated as follows with the diagrams shown in Fig.\ref{fig:mu_i}.

\begin{align*}
    &\mu_0=\text{Tr}(1)=1,\\
    &\mu_1=0,\\
    &\mu_2=J^2\text{Tr}\left[(\bm{S}_1\cdot\bm{S}_2)^2\right]=J^2\text{Tr}\left(S_{2z}^2\right)=\frac{1}{3}J^2,\\
    &\mu_3=0,\\
    &\mu_4^a=4!J^4\text{Tr}[(\bm{S}_1\cdot\bm{S}_2)(\bm{S}_2\cdot\bm{S}_3)(\bm{S}_3\cdot\bm{S}_4)(\bm{S}_4\cdot\bm{S}_1)]
    =24J^4\text{Tr}\left(S_{2z}^2 S_{3z}^2 S_{4z}^2\right)=\frac{8}{9}J^4,\\
    &\mu_4^b=J^4\text{Tr}\left[(\bm{S}_1\cdot\bm{S}_2)^4\right]=J^4\text{Tr}\left(S_{2z}^4\right)=\frac{1}{5}J^4\\
    &\mu_4^c=\frac{4!}{2!2!}\mu_2^2=\frac{2}{3}J^4.
\end{align*}
Since the terms in the traces above are all rotaionally invariant, we let one spin point to $z$ direction, 
which means $S_{1z}=1,S_{1x}=0,S_{1y}=0$ and simplifies the calculations.
This strategy is used in the whole section.

\subsection{Calculation of $\langle\rho_{11}^2\rangle$}
According to \eqref{equ:rho_ij}, 
\begin{equation}
    \langle\rho_{11}^2\rangle=\langle\rho_{11r}^2\rangle+2\langle\rho_{11r}\rho_{11l}\rangle+\langle\rho_{11l}^2\rangle
    =2\langle\rho_{11r}^2\rangle+2\langle\rho_{11r}\rho_{11l}\rangle,
\end{equation}
we will calculate $\langle\rho_{11r}^2\rangle$ and $\langle\rho_{11r}\rho_{11l}\rangle$ separately.

\subsubsection{Calculation of $\langle\rho_{11r}^2\rangle$}
\begin{figure}[htbp]
    \centering
    \includegraphics[scale=0.5]{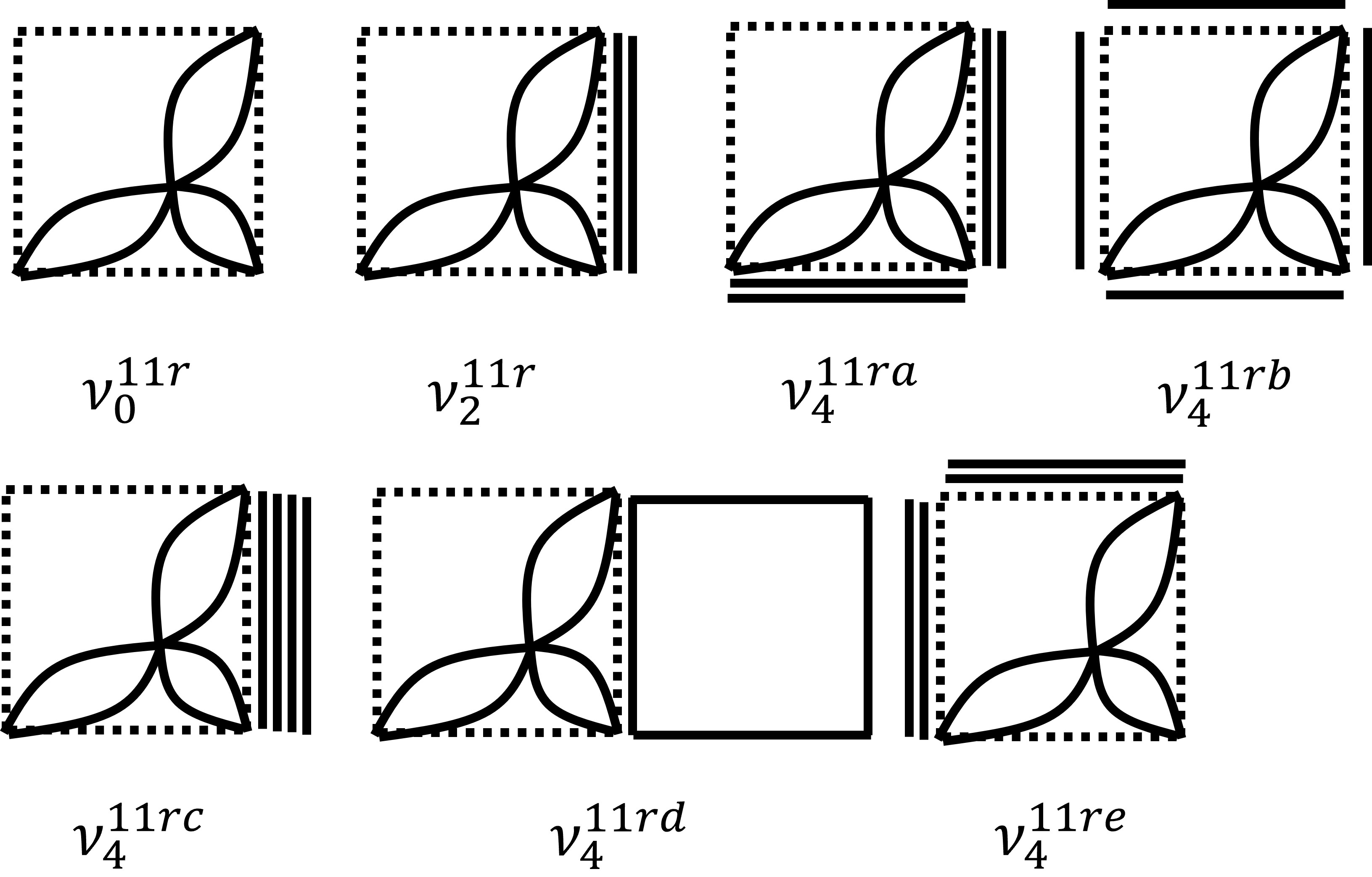}
    \caption{The diagrams for the first several nonzero $\nu_i$'s in $\rho_{11r}^2$}\label{fig:nu_11r}
\end{figure}

In order to calculate $\langle\rho_{11r}^2\rangle$, $\alpha_i$'s and $\nu_i$'s should be calculated first.
The rules~\cite{Stanley-1967a} for the diagrams of $\nu_i$'s are
\begin{itemize}
    \item Each site shoulde be connected with even number of lines.
    \item Only connected closed diagrams should be included.
    \item Diagrams connected on only one site should be excluded.
\end{itemize}
The diagrams for the first several nonzero $\nu_i$'s in $\rho_{11r}^2$ is shown in Fig.\ref{fig:nu_11r}.
\begin{align*}
    \nu_{0}^{11r}&=\text{Tr}\{[\bm{S}_{11}\cdot(\bm{S}_{21}\times\bm{S}_{22})]^2\}
    =\text{Tr}\left(S_{21y}^2 S_{22x}^2+S_{21x}^2 S_{22y}^2\right)=\frac{2}{9}\\
    \nu_{2}^{11r}&=J^2\text{Tr}\{[\bm{S}_{11}\cdot(\bm{S}_{21}\times\bm{S}_{22})]^2(\bm{S}_{21}\cdot\bm{S}_{22})^2\}
    =J^2\text{Tr}[(S_{11y}^2 S_{22x}^2+S_{11x}^2 S_{22y}^2)S_{22z}^2]=\frac{2}{45}J^2\\
    \nu_{4}^{11ra}&=\binom{4}{2}J^4\text{Tr}\{[\bm{S}_{11}\cdot(\bm{S}_{21}\times\bm{S}_{22})]^2(\bm{S}_{21}\cdot\bm{S}_{11})^2(\bm{S}_{21}\cdot\bm{S}_{22})^2\}\\
    &=6J^4\text{Tr}\left[(S_{11y}^2 S_{22x}^2+S_{11x}^2 S_{22y}^2) S_{11z}^2 S_{22z}^2\right]=\frac{4}{75}J^4\\
    \nu_{4}^{11rb}&=4!J^4\text{Tr}\{[\bm{S}_{11}\cdot(\bm{S}_{21}\times\bm{S}_{22})]^2(\bm{S}_{21}\cdot\bm{S}_{11})(\bm{S}_{21}\cdot\bm{S}_{22})(\bm{S}_{12}\cdot\bm{S}_{11})(\bm{S}_{12}\cdot\bm{S}_{22})\}\\
    &=24J^4\text{Tr}\left[(S_{11y}^2 S_{22x}^2+S_{11x}^2 S_{22y}^2) S_{11z}^2 S_{22z}^2S_{12z}^2\right]=\frac{16}{225}J^4\\
    \nu_{4}^{11rc}&=J^4\text{Tr}\{[\bm{S}_{11}\cdot(\bm{S}_{21}\times\bm{S}_{22})]^2(\bm{S}_{21}\cdot\bm{S}_{22})^4\}
    =J^4\text{Tr}[(S_{11y}^2 S_{22x}^2+S_{11x}^2 S_{22y}^2)S_{22z}^4]=\frac{2}{105}J^4\\
    \nu_{4}^{11rd}&=4!J^4\text{Tr}\{[\bm{S}_{11}\cdot(\bm{S}_{21}\times\bm{S}_{22})]^2(\bm{S}_{21}\cdot\bm{S}_{31})(\bm{S}_{21}\cdot\bm{S}_{22})(\bm{S}_{22}\cdot\bm{S}_{32})(\bm{S}_{32}\cdot\bm{S}_{31})\}\\
    &=24J^4\text{Tr}\left[(S_{11y}^2 S_{22x}^2+S_{11x}^2 S_{22y}^2) S_{22z}^2 S_{31z}^2S_{32z}^2\right]=\frac{16}{135}J^4\\
    \nu_{4}^{11re}&=\binom{4}{2}J^4\text{Tr}\{[\bm{S}_{11}\cdot(\bm{S}_{21}\times\bm{S}_{22})]^2(\bm{S}_{12}\cdot\bm{S}_{11})^2(\bm{S}_{12}\cdot\bm{S}_{22})^2\}\\
    &=6J^4\text{Tr}\left[(S_{21y}^2 S_{22x}^2+S_{21x}^2 S_{22y}^2) S_{12z}^2(S_{12x}^2S_{22x}^2+S_{12y}^2S_{22y}^2+S_{12z}^2S_{22z}^2)\right]\\
    &=24J^4\text{Tr}(S_{12x}^2 S_{12z}^2 S_{21y}^2 S_{22x}^4)+12J^4\text{Tr}(S_{12y}^2 S_{12z}^2 S_{21y}^2 S_{22x}^2 S_{22y}^2)
    =\frac{28}{225}J^4
\end{align*}

Then we can calculate $\alpha_l$'s from \eqref{equ:alpha_l}.
\begin{align*}
    \alpha_0^{11r}&=\nu_0^{11r}=\frac{2}{9},\\
    \alpha_1^{11r}&=0,\\
    \alpha_2^{11r}&=\nu_2^{11r}-\alpha_0^{11r}\mu_2=-\frac{4}{135}J^2,\\
    \alpha_3^{11r}&=0,\\
    \alpha_4^{11ra}&=\nu_4^{11ra}-\binom{4}{2}\alpha_2^{11r}\mu_2 \times 2-\alpha_0^{11r}\mu_4^c=\frac{16}{675}J^4,\\
    \alpha_4^{11rb}&=\nu_4^{11rb}-\alpha_0^{11r}\mu_4^a=-\frac{256}{2025}J^4,\\
    \alpha_4^{11rc}&=\nu_4^{11rc}-\binom{4}{2}\alpha_2^{11r}\mu_2-\alpha_0^{11r}\mu_4^b=\frac{32}{945}J^4,\\
    \alpha_4^{11rd}&=\nu_4^{11rd}-\alpha_0^{11r}\mu_4^a=-\frac{32}{405}J^4,\\
    \alpha_4^{11re}&=\nu_4^{11rd}-\alpha_0^{11r}\mu_4^c=-\frac{16}{675}J^4.
\end{align*}
Consequently, according to \eqref{equ:O_expansion}, $\langle\rho_{11r}^2\rangle$ can be expanded as
\begin{align}
    \begin{split}
        \langle\rho_{11r}^2\rangle=&\alpha_0+\frac{(-1)^2}{2!}\beta^2\alpha_2^{11r}\cdot 2
        +\frac{(-1)^4}{4!}\beta^4\left(\alpha_4^{11ra}+\alpha_4^{11rb}+2\alpha_4^{11rc}+2\alpha_4^{11rd}+\alpha_r^{11re}\right)+o(\beta^4)\\
        =&\frac{2}{9}-\frac{4}{135}\beta^2 J^2-\frac{128}{14175}\beta^4 J^4+o(\beta^4)
    \end{split}
\end{align}

\subsubsection{Calculation of $\langle\rho_{11r}\rho_{11l}\rangle$}\label{subsub:rho_11rrho_11l}
\begin{figure}[htbp]
    \centering
    \includegraphics[scale=0.5]{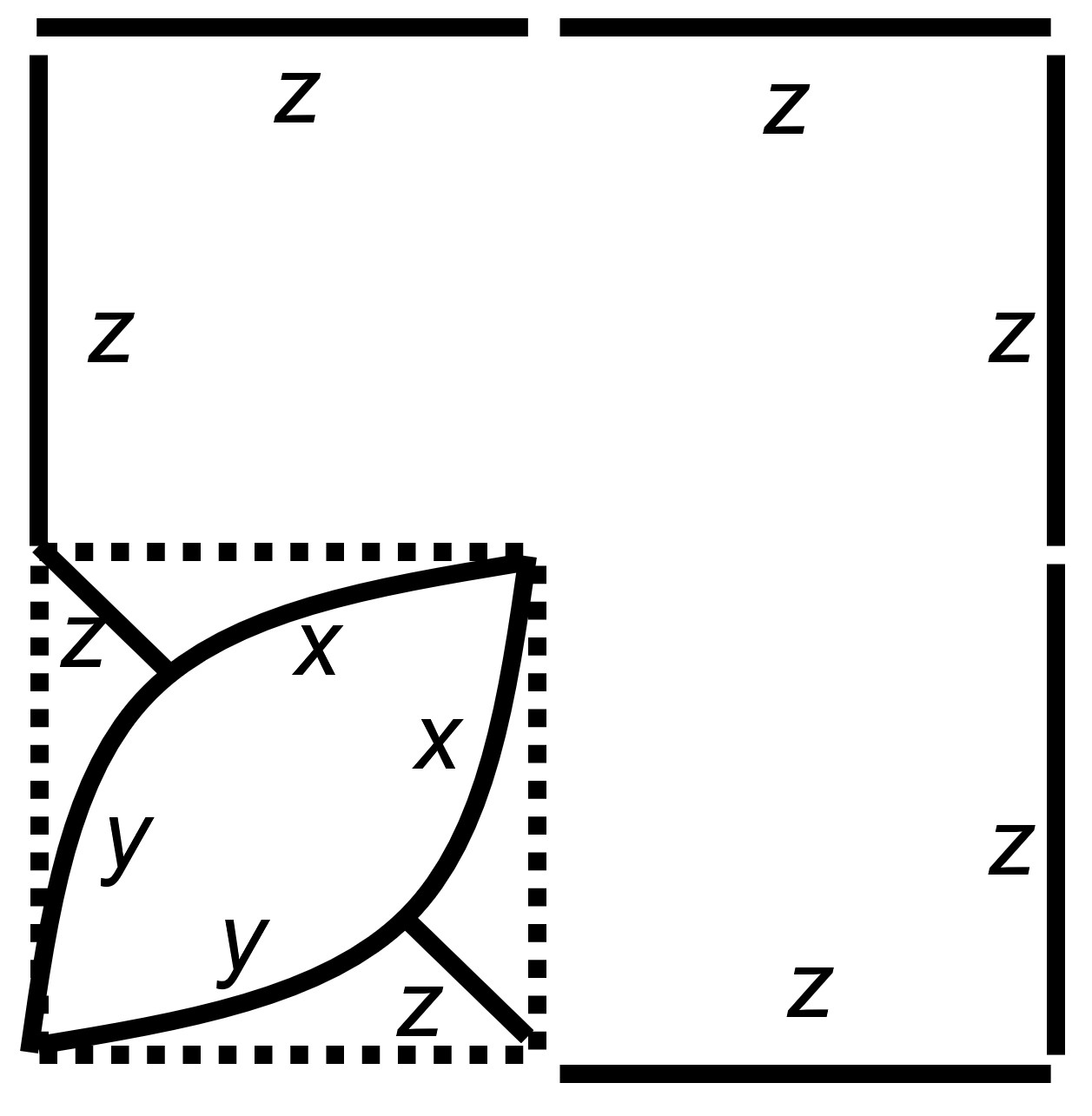}
    \caption{ The diagrams for the first nonzero $\nu_i$ in $\rho_{11r}\rho_{11l}$}\label{fig:nu_11rl}
\end{figure}

In order to identify the diagrams that contribute to $\langle\rho_{11r}\rho_{11l}\rangle$, each bond is labelled with $x$, $y$ or $z$,
which represents the spin components of the two sites linked by the bond. (see Fig.\ref{fig:nu_11rl})
The three bonds relating to $\rho_{11r}$ or $\rho_{11l}$ should be labelled with $x$, $y$ and $z$ respectively, due to the triple product.
The triple product also leads to another rule: 
\begin{itemize}
    \item The sites connected by the three bonds of one $\rho$ should not be connected by the bonds representing $\bm{S}_i\cdot\bm{S}_j$.
\end{itemize}
Therefore, the lowest order is six, which is represented by the diagram in Fig.\ref{fig:nu_11rl}.

\subsubsection{Summary of This Subsection}
In the end, $\langle \rho_{11}^2 \rangle$ can be calculated up to the fourth order.
\begin{equation}
    \langle\rho_{11}^2\rangle
    =2\langle\rho_{11r}^2\rangle+o(\beta^4)
    =\frac{4}{9}-\frac{8}{135}\beta^2 J^2-\frac{256}{14175}\beta^4 J^4+o(\beta^4).\label{eqn:rho_11^2}
\end{equation}

\subsection{Calculation of $\langle \rho_{11}\rho_{21} \rangle$}
\begin{figure}[htbp]
    \centering
    \includegraphics[scale=0.4]{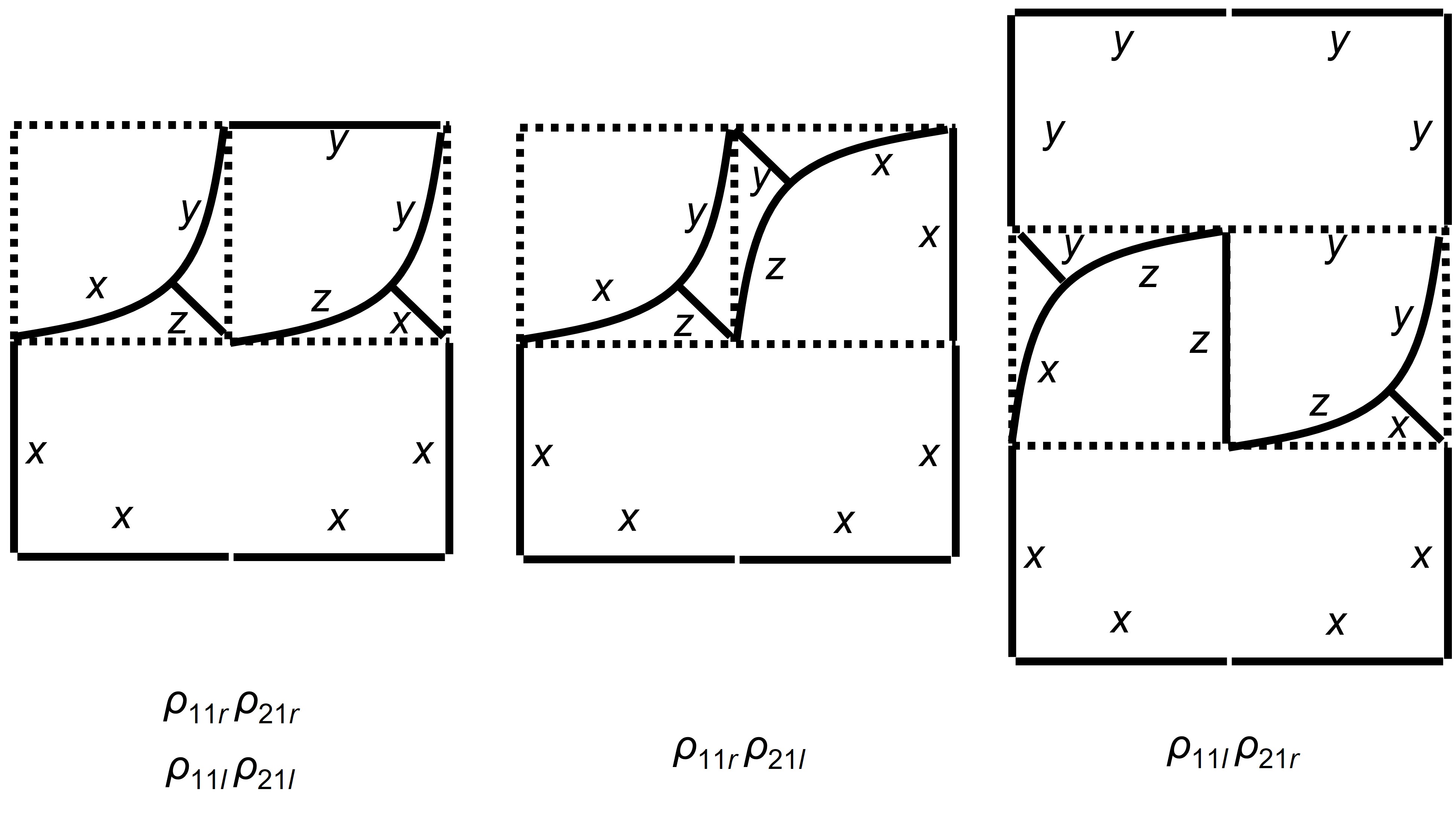}
    \caption{ The diagrams for the first several nonzero $\nu_i$'s in $\rho_{11}\rho_{21}$}\label{fig:rho_11rho_21}
\end{figure}

According to \eqref{equ:rho_ij}, 
\begin{equation}
    \langle \rho_{11}\rho_{21} \rangle=\langle \rho_{11r}\rho_{21r} \rangle+\langle \rho_{11r}\rho_{21l} \rangle
    +\langle \rho_{11l}\rho_{21r} \rangle+\langle \rho_{11l}\rho_{21l} \rangle.
\end{equation}
In order to follow the rule proposed in \ref{subsub:rho_11rrho_11l}, 
the diagrams for the first several nonzero $\nu_i$'s are sketched in Figure \ref{fig:rho_11rho_21}.

The lowest order for $\rho_{11r}\rho_{21r}$, $\rho_{11l}\rho_{21l}$ and $\rho_{11r}\rho_{21l}$ is 5,
while it is 9 for $\rho_{11l}\rho_{21r}$. As a result, 
\begin{equation}
    \langle \rho_{11}\rho_{21} \rangle=0+o(\beta^4).\label{eqn:rho_11rho_21}
\end{equation}

\subsection{Calculation of $\langle \rho_{11}\rho_{22} \rangle$}
\begin{figure}[htbp]
    \centering
    \includegraphics[scale=0.4]{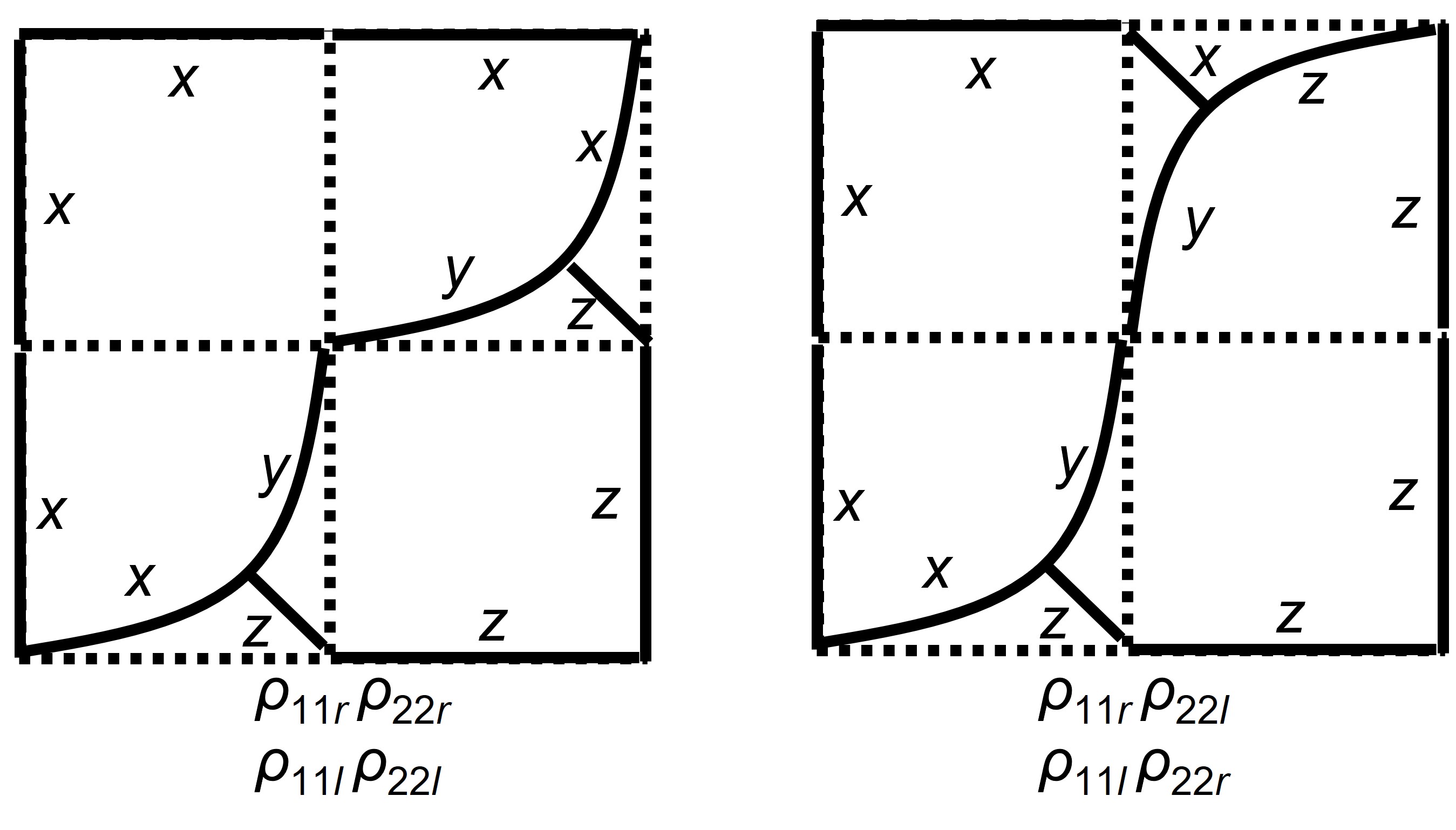}
    \caption{ The diagrams for the first several nonzero $\nu_i$'s in $\rho_{11}\rho_{22}$}\label{fig:rho_11rho_22}
\end{figure}

According to \eqref{equ:rho_ij}, 
\begin{equation}
    \langle \rho_{11}\rho_{22} \rangle=\langle \rho_{11r}\rho_{22r} \rangle+\langle \rho_{11r}\rho_{22l} \rangle
    +\langle \rho_{11l}\rho_{22r} \rangle+\langle \rho_{11l}\rho_{22l} \rangle.
\end{equation}
In order to follow the rule proposed in \ref{subsub:rho_11rrho_11l}, 
the diagrams for the first several nonzero $\nu_i$'s are sketched in Figure \ref{fig:rho_11rho_22}.

The lowest order for each term is 6. As a result, 
\begin{equation}
    \langle \rho_{11}\rho_{22} \rangle=0+o(\beta^4).\label{eqn:rho_11rho_22}
\end{equation}

\subsection{Calculation of Calculation of $\langle\rho_{12}\rho_{21}\rangle$}
\begin{figure}[htbp]
    \centering
    \includegraphics[scale=0.5]{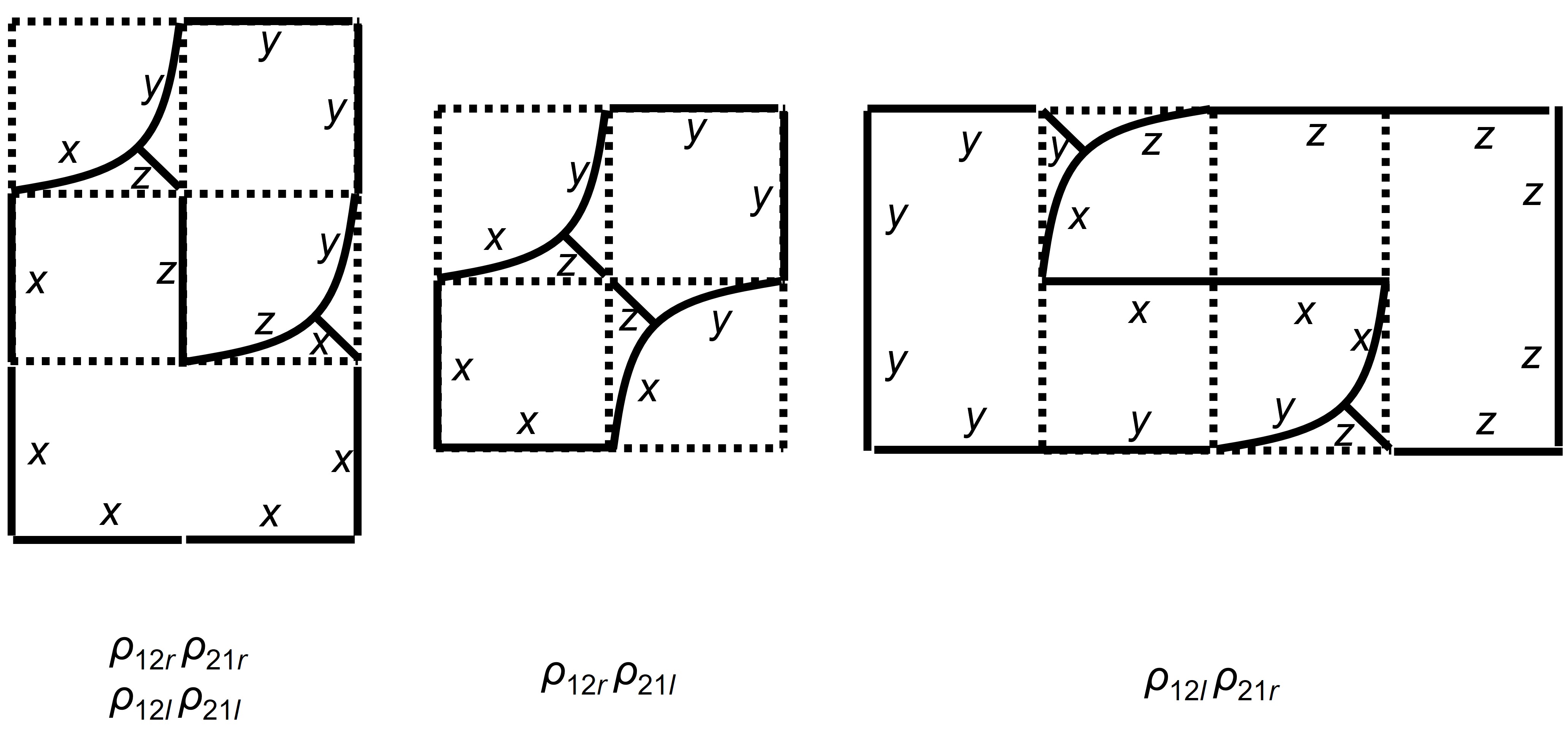}
    \caption{ The diagrams for the first several nonzero $\nu_i$'s in $\rho_{12}\rho_{21}$}\label{fig:rho_12rho_21}
\end{figure}

According to \eqref{equ:rho_ij}, 
\begin{equation}
    \langle \rho_{12}\rho_{21} \rangle=\langle \rho_{12r}\rho_{21r} \rangle+\langle \rho_{12r}\rho_{21l} \rangle
    +\langle \rho_{12l}\rho_{21r} \rangle+\langle \rho_{12l}\rho_{21l} \rangle.
\end{equation}
In order to follow the rule proposed in \ref{subsub:rho_11rrho_11l}, 
the diagrams for the first several nonzero $\nu_i$'s are sketched in Figure \ref{fig:rho_12rho_21}.

The lowest order is 8 for $\rho_{12r}\rho_{21r}$ or $\rho_{12l}\rho_{21l}$ and 12 for $\rho_{12l}\rho_{21r}$, 
while it is 4 for $\rho_{12r}\rho_{21l}$, which should be taken into account in the summation.

\begin{equation*}
    \alpha_4^{12r21l}=\nu_4^{12r21l}=(-1)\cdot 2\cdot 4!J^4\cdot
    \text{Tr}\left(S_{11x}^2 S_{21x}^2 S_{12x}^2 S_{32y}^2 S_{33y}^2 S_{23y}^2\right)
    =-\frac{16}{243}J^4.
\end{equation*}
Therefore, 
\begin{equation}
    \langle \rho_{12}\rho_{21} \rangle=\frac{(-1)^4}{4!}\beta^4 \alpha_4^{12r21l}+o(\beta^4)=-\frac{2}{729}\beta^4 J^4+o(\beta^4)\label{eqn:rho_12rho_21}
\end{equation}

\subsection{Summary of This Section}
Thanks to the above subsections, the fluctuation of the topological charges can be calculated up to the fourth order from high-temperature expansion. Substitute \eqref{eqn:rho_11^2}, \eqref{eqn:rho_11rho_21}, \eqref{eqn:rho_11rho_22} and \eqref{eqn:rho_12rho_21} into \eqref{eqn:chi_L},
\begin{align}
    \begin{split}
        \chi_L=L^2\left(\frac{4}{9}-\frac{8}{135}\beta^2 J^2-\frac{256}{14175}\beta^4 J^4\right)
        +2(L-1)^2\left(-\frac{2}{729}\beta^4 J^4\right)+o(\beta^4)
    \end{split}
\end{align}
If the fluctuation is truncated at the second order, it is proportional to $L^2$.
Since area of the $L \times L$ square is proportional to $L^2$, the fluctuation of topological charges is proportional to the area at high temperatures.

\section{Low-temperature expansion}
At low temperatures, the spins will fluctuate around a certain direction.
Without loss of generality, the direction is along the $z$ axis.
Consequently, $\bm{S}=(m_x,m_y,\sqrt{1-m_x^2-m_y^2})$, where $m_x\ll 1,m_y\ll 1$.~\cite{Zinn-JustinBrezin-1976B}
In the continuous limit, the Hamiltonian is
\begin{align}
    \mathcal{H}=\frac{1}{2}J\iint(\partial_i \bm{S})\cdot(\partial_i \bm{S}) \dd\sigma
    =\frac{1}{2}J\iint\left[(\bm{\nabla}m_x)^2+(\bm{\nabla}m_y)^2\right]\dd\sigma,
\end{align}
where the terms with higher orders are omitted at low temperatures.
The partition function is
\begin{equation}
    Z_0=\int\mathcal{D}\bm{S}\left[\prod_x\delta(1-\bm{S}^2)\right]\exp(-\beta\mathcal{H})
    =\int\frac{\mathcal{D}\bm{m}}{\prod_x\sqrt{1-|\bm{m}|^2}}\exp(-\beta\mathcal{H})
    =\int\mathcal{D}\bm{m}\exp(-\beta\mathcal{H}),
\end{equation}
where the terms from the measure are omitted since they are of higher orders.
This partition function represents two decoupled massless free scalar fields.

In the continuous and low-temperature limit, the topological charge of a region $\Omega$ is
\begin{align}
    \begin{split}
        Q=&\iint_\Omega \bm{S}\cdot\left(\frac{\partial\bm{S}}{\partial x}\times\frac{\partial\bm{S}}{\partial y}\right)\dd \sigma\\
        =&\iint_\Omega \left(\frac{\partial m_x}{\partial x}\frac{\partial m_y}{\partial y}-\frac{\partial m_x}{\partial y}\frac{\partial m_y}{\partial x}\right)\dd \sigma\\
        =&\iint_\Omega \left[\frac{\partial}{\partial x}\left(m_x\frac{\partial m_y}{\partial y}\right)-\frac{\partial}{\partial y}\left(m_x\frac{\partial m_y}{\partial x}\right)\right]\dd \sigma.
    \end{split}
\end{align}
To simplify it, we define $\bm{A}=(m_x\partial_x m_y,m_x\partial_y m_y)=(A_x,A_y)$, and through Green's theorem the above expression becomes
\begin{align}
    Q=\iint_\Omega \left(\frac{\partial A_y}{\partial x}-\frac{\partial A_x}{\partial y}\right)\dd \sigma
    =\oint_{\partial\Omega} \left(A_x \,\dd x + A_y \,\dd y\right),
\end{align}
where $\partial\Omega$ represents the boundary of the region $\Omega$.

The region is chosen to be a circle for simplicity. Its centre is situated at the origin of the coordinate system and its radius is $L$.
Considering the symmetry, we can also transform the Cartesian coordinates $(x,y)$ into the polar coordinates $(\rho,\theta)$,
which means 
\begin{align*}
    &A_x=m_x\left[\frac{\partial m_y}{\partial\rho}\cos\theta+\frac{\partial m_y}{\partial\theta}\left(-\frac{\sin\theta}{\rho}\right)\right],\\
    &A_y=m_x\left[\frac{\partial m_y}{\partial\rho}\sin\theta+\frac{\partial m_y}{\partial\theta}\left(\frac{\cos\theta}{\rho}\right)\right],\\
    &\dd x=-\rho\sin\theta \dd \theta,\\
    &\dd y=\rho\cos\theta \dd \theta.
\end{align*}
Therefore,
\begin{equation}
    Q=\int_0^{2\pi}m_x(L,\theta) \frac{\dd m_y(L,\theta)}{\dd \theta}\dd \theta.
\end{equation}
Since $m_x$ and $m_y$ are decoupled, $\langle Q\rangle =0$. So the fluctuation of the topological charge in a circular region with radius $L$ is
\begin{equation}
    \chi_L=\langle Q^2 \rangle=\int_0^{2\pi}\dd\theta\int_{0}^{2\pi}\dd\theta'\langle m_x(L,\theta)m_x(L,\theta')\rangle \left\langle \frac{\dd m_y(L,\theta)}{\dd \theta}\frac{\dd m_y(L,\theta')}{\dd \theta'}\right\rangle
\end{equation}
According to appendix \ref{app:correlators}, 
\begin{equation}
    \langle m_x(L,\theta)m_x(L,\theta')\rangle=G_{xx}\left(2L\sin\frac{|\theta-\theta'|}{2}\right)
\end{equation}
Also, we can get
\begin{equation}
    \left\langle \frac{\dd m_y(L,\theta)}{\dd \theta}\frac{\dd m_y(L,\theta')}{\dd \theta'}\right\rangle
    =\frac{\partial}{\partial\theta}\frac{\partial}{\partial\theta'}G_{xx}\left(2L\sin\frac{|\theta-\theta'|}{2}\right)
    =-\frac{\partial^2}{\partial\theta^2}G_{xx}\left(2L\sin\frac{|\theta-\theta'|}{2}\right)
\end{equation}

\begin{figure}[H]
    \centering
    \includegraphics[scale=0.5]{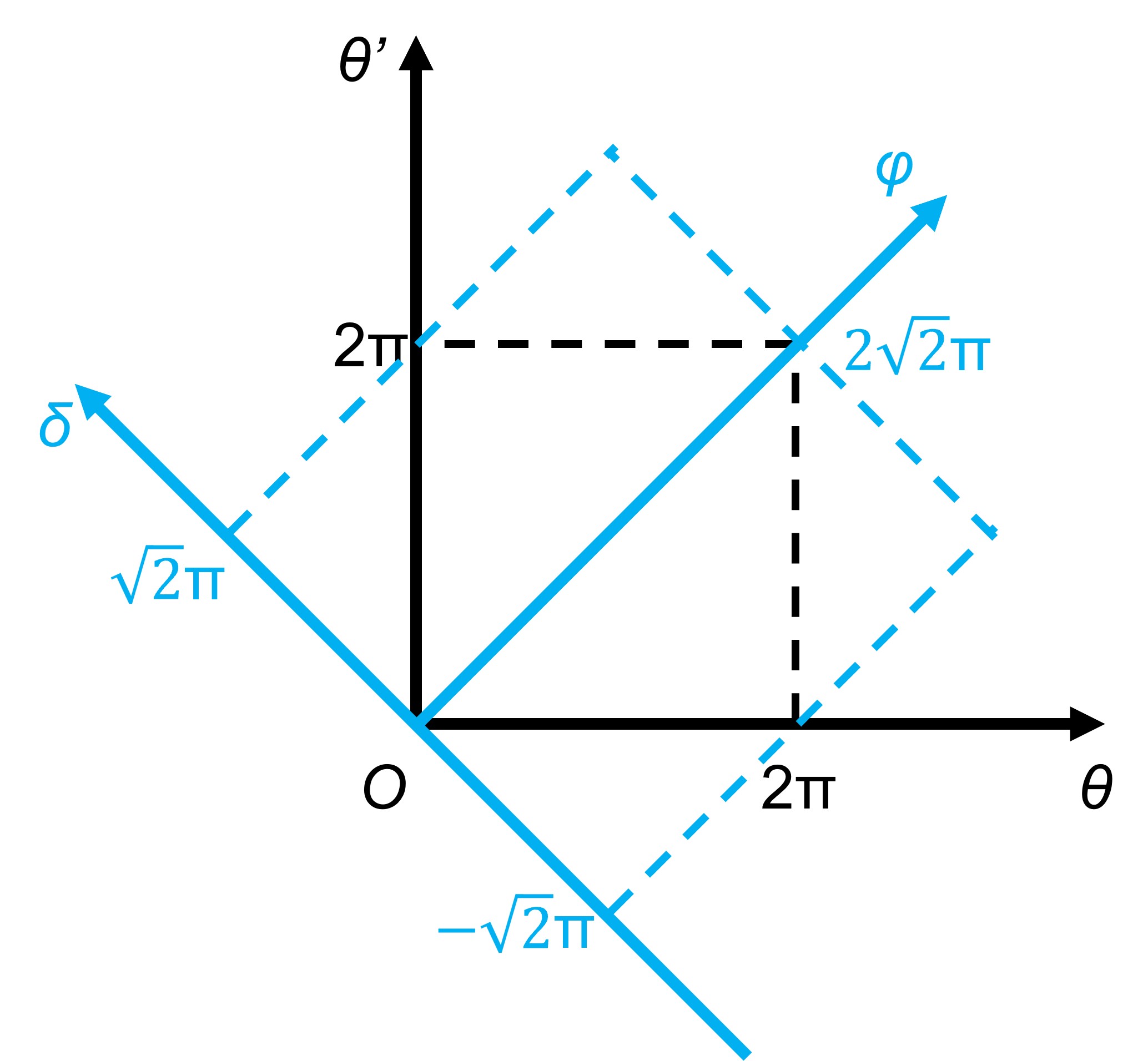}
    \caption{ The graph for the coordinate transformation }\label{fig:CoordinateTransform}
\end{figure}
For simplicity, we perform the coordinate transformation
\begin{equation}
    \delta=\frac{\theta'-\theta}{\sqrt{2}},\varphi=\frac{\theta'+\theta}{\sqrt{2}},
\end{equation}
and enlarge the integral region, which becomes the region enveloped by the blue dashed lines and the $\delta$ axis in Figure \ref{fig:CoordinateTransform}, due to the periodical integrand.
Therefore, the fluctuation becomes
\begin{align}
    \begin{split}
        \chi_L=&-\frac{1}{2}\int_0^{2\sqrt{2}\pi}\dd\varphi\int_{-\sqrt{2}\pi}^{\sqrt{2}\pi}\dd\delta G_{xx}\left(2L\sin\frac{|\delta|}{\sqrt{2}}\right)\left(\frac{1}{2}\frac{\dd^2}{\dd\delta^2}\right)G_{xx}\left(2L\sin\frac{|\delta|}{\sqrt{2}}\right)\\
        =&-\sqrt{2}\pi\int_{0}^{\sqrt{2}\pi}\dd\delta G_{xx}\left(2L\sin\frac{\delta}{\sqrt{2}}\right)\frac{\dd^2}{\dd\delta^2}G_{xx}\left(2L\sin\frac{\delta}{\sqrt{2}}\right)\\
        =&-2\sqrt{2}\pi\int_{0}^{\frac{\sqrt{2}}{2}\pi}\dd\delta G_{xx}\left(2L\sin\frac{\delta}{\sqrt{2}}\right)\frac{\dd^2}{\dd\delta^2}G_{xx}\left(2L\sin\frac{\delta}{\sqrt{2}}\right)\\
        =&-2\sqrt{2}\pi\left[G_{xx}\left(2L\sin\frac{\delta}{\sqrt{2}}\right)\frac{\dd}{\dd\delta}G_{xx}\left(2L\sin\frac{\delta}{\sqrt{2}}\right)\right]_0^{\frac{\sqrt{2}}{2}\pi}\\
        &+2\sqrt{2}\pi\int_{0}^{\frac{\sqrt{2}}{2}\pi}\dd\delta\left[\frac{\dd}{\dd\delta}G_{xx}\left(2L\sin\frac{\delta}{\sqrt{2}}\right)\right]^2.\label{equ:chiL_Gxx}
    \end{split}
\end{align}
According to Appendix \ref{app:correlators}, 
\begin{align}
    \begin{split}
        \frac{\dd}{\dd\delta}G_{xx}\left(2L\sin\frac{\delta}{\sqrt{2}}\right)=&\sqrt{2}L\cos\frac{\delta}{\sqrt{2}}G_{xx}'\left(2L\sin\frac{\delta}{\sqrt{2}}\right)
    \end{split}
\end{align}
Since $\cos\frac{\pi}{2}=0$ and $G_{xx}'(0)=0$, the first term in \eqref{equ:chiL_Gxx} is zero. Consequently,
\begin{align}
    \begin{split}
        \chi_L=&4\sqrt{2}\pi L^2 \int_{0}^{\frac{\sqrt{2}}{2}\pi}\dd\delta \cos^2\left(\frac{\delta}{\sqrt{2}}\right)G_{xx}'^2\left(2L\sin\frac{\delta}{\sqrt{2}}\right)\\
        =&8\pi L^2\int_{0}^{\frac{\pi}{2}}\dd\alpha \cos^2\alpha G_{xx}'^2\left(2L\sin\alpha\right)\\
        =&\frac{1}{2\pi\beta^2 J^2}\int_{0}^{\frac{\pi}{2}}\dd\alpha \cos^2\alpha \left[\frac{J_0\left(\frac{2L}{a}\sin\alpha\right)-1}{\sin\alpha}\right]^2\\
        =&\frac{L}{2\pi a \beta^2 J^2}\int_{0}^{\frac{L}{a}}\dd v \sqrt{1-\left(\frac{va}{L}\right)^2} \left[\frac{J_0(2v)-1}{v}\right]^2\\
        =&\frac{L}{2\pi a \beta^2 J^2}\int_{0}^{\frac{L}{a}}\dd v \left[1-\frac{1}{2}\left(\frac{a}{L}\right)^2 v^2-\frac{1}{8}\left(\frac{a}{L}\right)^4 v^4-\frac{1}{16}\left(\frac{a}{L}\right)^6 v^6+o(v^6)\right]\left[\frac{J_0(2v)-1}{v}\right]^2,
    \end{split}
\end{align}
where $v=\frac{L}{a}\sin\alpha$ is substituted in the third line.
It is clear that only the first term contribute in the limit $L\rightarrow\infty$, since the other terms will approach a constant. Due to the decaying behaviour of $\left[\frac{J_0(2v)-1}{v}\right]^2$ for a large $v$, the upper limit of the integral can be approximated by infinity. Therefore,
\begin{equation}
    \chi_L\approx\frac{L}{2\pi a \beta^2 J^2}\int_{0}^{\infty}\dd v \left[\frac{J_0(2v)-1}{v}\right]^2\propto L,
\end{equation}
which implies that the fluctuation of the topolotical charges is proportional to $L$. Since the perimeter of the circle is proportional to $L$, the fluctuation of topological charges is proportional to the perimeter at low temperatures.

\section{Conclusion}
We have calculated the fluctuation of the topological charges encircled by a region in the two-dimensional classical Heisenberg model theoretically from high-temperature expansion and low-temperature expansion. 
At high temperatures, the fluctuation of the topological charges on a discrete lattice is calculated by being expanded at the orders of $\beta=\frac{1}{k_B T}$. Through a lot of diagrams, we have found that the fluctuation is proportional to the area of the region at high temperatures.
At low temperatures, the fluctuation of the topological charges in the continuous limit is calculated by being expanded at the orders of $\frac{1}{\beta}=k_B T$. With the help of the correlation functions, we have found that the flucation is proportional to the perimieter of the region at low temperatures.
In the end, we have proven the simulation results in our previous study that the fluctuation of the topological charges satisfies the area law at high temperatures while complying with the perimeter law at low temperatures. 
These results strongly suggest that there is a transition in between in the two-dimensional classical Heisenberg model, just like the Kosterlitz-Thouless transition in the XY model.
However, the transition temperature has not been calculated yet, which calls for further researches.

\newpage
\appendix
\section{Calculation of Tr$\left(S_x^{2k} S_y^{2l} S_z^{2m}\right)$}\label{app:Tr_SxSySz}
In the spherical coordinate, $S_x=\sin\theta \cos\phi$, $S_y=\sin\theta \sin\phi$, $S_z=\cos\theta$. 
Therefore,
\begin{align}
    \begin{split}
        \text{Tr}\left(S_x^{2k} S_y^{2l} S_z^{2m}\right)
        =&\frac{1}{4\pi}\iint (\sin\theta \cos\phi)^{2k} (\sin\theta \sin\phi)^{2l} (\cos\theta)^{2m} \sin\theta \dd\theta \dd\phi\\
        =&\frac{1}{4\pi}\left[\int_0^\pi (\sin\theta)^{2k+2l+1} (\cos\theta)^{2m} \dd\theta\right]
        \left[\int_0^{2\pi} (\cos\phi)^{2k} (\sin\phi)^{2m} \dd\phi\right]
    \end{split}
\end{align}
Since $B(x,y)=2\int_0^1 t^{2x-1} (1-t^2)^{y-1} \dd t$, the above expression can be transformed to
\begin{align}
    \begin{split}
        \text{Tr}\left(S_x^{2k} S_y^{2l} S_z^{2m}\right)
        =&\frac{1}{4\pi}B\left(m+\frac{1}{2},k+l+1\right)\cdot 2B\left(k+\frac{1}{2},l+\frac{1}{2}\right)\\
        =&\frac{1}{2\pi}\frac{\Gamma\left(k+\frac{1}{2}\right)\Gamma\left(l+\frac{1}{2}\right)\Gamma\left(m+\frac{1}{2}\right)}{\Gamma\left(k+l+m+\frac{3}{2}\right)},
    \end{split}
\end{align}
with substitution $t_1=\cos\theta,t_2=\sin\phi$.

\section{Correlation functions}\label{app:correlators}
Since $m_x$ and $m_y$ are decoupled, $Z_0=Z_x Z_y$, where
\begin{equation}
    Z_x=\int \mathcal{D}m_x \exp\left[-\frac{1}{2}\beta J\iint(\bm{\nabla}m_x)^2 \dd \sigma\right]
    =\int \mathcal{D}m_x \exp\left[-\frac{1}{2}\beta J\iint m_x (-\bm{\nabla}^2 m_x) \dd \sigma\right],
\end{equation}
and the expression for $Z_y$ is similar except that the subscripts are replaced by $y$.
In order to calculate the correlation functions, we introduce a local source function $H(\bm{r})$, where $\bm{r}=(x,y)$.
Therefore, the partition function becomes
\begin{align}
    \begin{split}
        Z_x(\{H\})=&\int \mathcal{D}m_x \exp\left[-\frac{1}{2}\beta J\iint m_x (-\bm{\nabla}^2 m_x) \dd \sigma
        +\iint m_x(\bm{r})H(\bm{r}) \dd \sigma\right]\\
        =&\int \mathcal{D}m_x \exp\left\{-\frac{1}{2}\iint m_x(\bm{r}) \left[-\beta J\delta(\bm{r}-\bm{r}')\bm{\nabla}'^2\right] m_x(\bm{r}')\dd \sigma \dd \sigma'
        +\iint m_x(\bm{r})H(\bm{r}) \dd \sigma\right\}.
    \end{split}
\end{align}
Then, we can define an operator
\begin{equation}
    \hat{\mathcal{O}}(\bm{r}-\bm{r}')=-\beta J\delta(\bm{r}-\bm{r}')\bm{\nabla}'^2.
\end{equation}
In order to simplify the functional integration, $m_x$ can be substituted for 
\begin{equation}
    m_x(\bm{r})=m_x^*(\bm{r})+\iint \hat{\mathcal{O}}^{-1}(\bm{r}-\bm{r}')H(\bm{r}')\dd \sigma',
\end{equation}
where the inverse operator is defined as $\iint\hat{\mathcal{O}}(\bm{r}-\bm{r}'')\hat{\mathcal{O}}^{-1}(\bm{r}''-\bm{r}') \dd \sigma''=\delta(\bm{r}-\bm{r}')$.
As a result,
\begin{align}
    \begin{split}
        Z_x(\{H\})=&\int \mathcal{D}m_x^* \exp\left\{-\frac{1}{2}\iint \left[m_x^*(\bm{r})+\iint \hat{\mathcal{O}}^{-1}(\bm{r}-\bm{r}'')H(\bm{r}'')\dd \sigma''\right]\hat{\mathcal{O}}(\bm{r}-\bm{r}')\right.\\
        &\times \left[m_x^*(\bm{r}')+\iint \hat{\mathcal{O}}^{-1}(\bm{r}'-\bm{r}''')H(\bm{r}''')\dd \sigma'''\right]\dd \sigma \dd \sigma'\\
        &\left.+\iint \left[m_x^*(\bm{r})+\iint \hat{\mathcal{O}}^{-1}(\bm{r}-\bm{r}')H(\bm{r}')\dd \sigma'\right]H(\bm{r}) \dd \sigma\right\}\\
        =&\int \mathcal{D}m_x^* \exp\left\{-\frac{1}{2}\iint m_x^*(\bm{r})\hat{\mathcal{O}}(\bm{r}-\bm{r}')m_x^*(\bm{r}') \dd\sigma \dd\sigma'\right.\\
        &\left.+\frac{1}{2}\iint H(\bm{r})\hat{\mathcal{O}}^{-1}(\bm{r}-\bm{r}')H(\bm{r}')\dd\sigma \dd \sigma'\right\}\\
        =&Z_x(0)\exp\left[\frac{1}{2}\iint H(\bm{r})\hat{\mathcal{O}}^{-1}(\bm{r}-\bm{r}')H(\bm{r}')\dd\sigma \dd \sigma'\right].
    \end{split}
\end{align}
Then we can obtain the correlation functions
\begin{equation}
    \langle m_x(\bm{r})m_x(\bm{r}') \rangle=\frac{1}{Z_x(0)}\left.\frac{\delta^2 Z_x(\{H\})}{\delta H(\bm{r})\delta H(\bm{r}')}\right|_{H=0}
    =\hat{\mathcal{O}}^{-1}(\bm{r}-\bm{r}') \label{eqn:correlationFunction}
\end{equation}
In order to calculate $\hat{\mathcal{O}}^{-1}(\bm{r}-\bm{r}')$, we can first calculate the Fourier transformation of it, which is defined as
\begin{align}
    &\hat{\mathcal{O}}(\bm{r}-\bm{r}')=\iint \frac{d^2\bm{k}}{(2\pi)^2}e^{i\bm{k}\cdot(\bm{r}-\bm{r}')}\hat{\mathcal{O}}(\bm{k}),\\
    &\hat{\mathcal{O}}^{-1}(\bm{r}-\bm{r}')=\iint \frac{d^2\bm{k}}{(2\pi)^2}e^{i\bm{k}\cdot(\bm{r}-\bm{r}')}\hat{\mathcal{O}}^{-1}(\bm{k}),\label{eqn:O-1rr0}
\end{align}
which satisfies $\hat{\mathcal{O}}(\bm{k})\hat{\mathcal{O}}^{-1}(\bm{k})=1$.
Since
\begin{equation}
    \hat{\mathcal{O}}(\bm{k})=\iint \hat{\mathcal{O}}(\bm{r}-\bm{r}')e^{-i\bm{k}\cdot(\bm{r}-\bm{r}')} d\sigma
    =\beta J \bm{k}^2, 
\end{equation}
we can get
\begin{equation}
    \hat{\mathcal{O}}^{-1}(\bm{k})=\frac{1}{\beta J \bm{k}^2},\label{eqn:O-1k}
\end{equation}
In order to avoid the infrared and ultraviolet divergence of the integral, an infrared cutoff $\frac{1}{Na}$ and an ultraviolet cutoff $\frac{1}{a}$ should be introduced, where $a$ is the lattice constant and $N$ is the linear size of the system.
Therefore, combing \eqref{eqn:correlationFunction}, \eqref{eqn:O-1rr0} and \eqref{eqn:O-1k}, we can get
\begin{align}
    \begin{split}
        \langle m_x(\bm{r})m_x(\bm{r}') \rangle\equiv G_{xx}(|\bm{r}-\bm{r'}|)=&\frac{1}{\beta J}\iint \frac{d^2\bm{k}}{(2\pi)^2}\frac{e^{i\bm{k}\cdot(\bm{r}-\bm{r}')}}{\bm{k}^2}\\
        =&\frac{1}{4\pi^2 \beta J}\int_\frac{1}{Na}^\frac{1}{a} \frac{k\dd k}{k^2}\int_0^{2\pi}e^{ik|\bm{r}-\bm{r}'|\cos\theta}\dd \theta\\
        =&\frac{1}{2\pi \beta J}\int_\frac{1}{Na}^\frac{1}{a} \frac{J_0(k|\bm{r}-\bm{r}'|)}{k}\dd k.\\
    \end{split}
\end{align}
where $J_0$ is the Bessel function of the first kind. Since $J_0$ can be expanded as
\begin{equation*}
    J_0(k|\bm{r}-\bm{r}'|)=\sum_{n=0}^{\infty}\frac{(-1)^n k^{2n}|\bm{r}-\bm{r'}|^{2n}}{2^{2n}(n!)^2},
\end{equation*}
we can obtain
\begin{align}
    \begin{split}
        G_{xx}(|\bm{r}-\bm{r}'|)=&\frac{1}{2\pi \beta J}\sum_{n=0}^{\infty}\frac{(-1)^n|\bm{r}-\bm{r'}|^{2n}}{2^{2n}(n!)^2}\int_\frac{1}{Na}^\frac{1}{a} \frac{k^{2n}}{k}\dd k\\
        =&\frac{1}{2\pi \beta J}\left[\ln N+\sum_{n=1}^{\infty}\frac{(-1)^n|\bm{r}-\bm{r'}|^{2n}}{2^{2n}(n!)^2}\frac{1}{2n}\left(\frac{1}{a^{2n}}-\frac{1}{N^{2n}a^{2n}}\right)\right]\\
        \approx&\frac{1}{2\pi \beta J}\left[\ln N+\sum_{n=1}^{\infty}\frac{(-1)^n|\bm{r}-\bm{r'}|^{2n}}{2n (2a)^{2n}(n!)^2}\right],
    \end{split}
\end{align}
where $N\gg  1$ is used. Since $\langle m_y(\bm{r})m_y(\bm{r}') \rangle$ possesses the same expression, they can be combined as
\begin{equation}
    \langle m_i(\bm{r})m_j(\bm{r}') \rangle=\delta_{ij}G_{xx}(|\bm{r}-\bm{r}'|).
\end{equation}
The derivative of $G_{xx}$ is
\begin{equation}
    G_{xx}'(|\bm{r}-\bm{r}'|)=\frac{1}{2\pi \beta J}\sum_{n=1}^{\infty}\frac{(-1)^n|\bm{r}-\bm{r'}|^{2n-1}}{(2a)^{2n}(n!)^2}
    =\frac{1}{2\pi \beta J}\frac{J_0\left(\frac{|\bm{r}-\bm{r'}|}{a}\right)-1}{|\bm{r}-\bm{r}'|}.
\end{equation}
Clearly, $G_{xx}'(0)=0$.

\end{document}